\documentclass[aps,pra,reprint,amsmath,amssymb,groupedaddress,showpacs]{revtex4-1}
\usepackage{graphicx}
\usepackage{float}
\usepackage{dcolumn} 
\usepackage{bm}      
\usepackage[bookmarks=false]{hyperref} 
\usepackage[font={small}]{caption} 
\usepackage{enumitem}
\usepackage{pgfplots}
\pgfplotsset{compat=1.9}
\hypersetup{
    unicode=false,          
    pdftoolbar=true,        
    pdfmenubar=true,        
    pdffitwindow=true,      
    pdfstartview={},        
    pdftitle={X States of the Same Spectrum and Entanglement as All Two-Qubit States}, 
    pdfauthor={Samuel R. Hedemann},
    pdfsubject={Quantum Entanglement},   
    pdfcreator={},          
    pdfproducer={},         
    pdfkeywords={X States,} {Entanglement-Preserving Unitary,} {EPU,} {EPU Equivalence,} {Entanglement,} {Concurrence,} {Maximally Entangled Mixed States,} {MEMS,} {True-Generalized X States,} {TGX States}, 
    pdfnewwindow=true,      
    colorlinks=true,        
    linkcolor=black,        
    citecolor=black,        
    filecolor=black,        
    urlcolor=black          
}
\newcommand{\Sec}[1]{\hyperref[sec:#1]{Sec.{\kern 2pt}\ref*{sec:#1}}}
\newcommand{\Section}[1]{\hyperref[sec:#1]{Section~\ref*{sec:#1}}}
\newcommand{\Fig}[2][]{\hyperref[fig:#2]{Fig.{\kern 2pt}\ref*{fig:#2}#1}}
\newcommand{\Figure}[2][]{\hyperref[fig:#2]{Figure~\ref*{fig:#2}#1}}
\newcommand{\App}[1]{\hyperref[sec:App.#1]{App.{\kern 2pt}\ref*{sec:App.#1}}}
\newcommand{\Appendix}[1]{\hyperref[sec:App.#1]{Appendix~\ref*{sec:App.#1}}}
\newcommand{\EqLab}[1]{\\\noindent\smash{\raisebox{6pt}[0pt][0pt]{\hypertarget{eq:#1}{}}}\vspace{-11pt}}
\newcommand{\Eq}[1]{\protect\hyperlink{eq:#1}{(\ref*{eq.#1})}}
\newcommand{\Eqs}[2]{\protect\hyperlink{eq:#1}{(\ref*{eq.#1}--\ref*{eq.#2})}}
\newenvironment{Equation}[1]{\EqLab{#1}\begin{equation}\label{eq.#1}}{\end{equation}\par\noindent\ignorespacesafterend}
\newcommand{\Table}[2][]{\hyperref[tab:#2]{Table~\ref*{tab:#2}#1}}
\newcommand{\ThmLab}[1]{\noindent\smash{\raisebox{11pt}[0pt][0pt]{\hypertarget{Thm:#1}{}}}\textbf{Theorem{\kern 3pt}#1:~~}}
\newcommand{\Thm}[1]{\protect\hyperlink{Thm:#1}{Theorem{\kern 3pt}#1}}
\newcommand{\Thms}[2]{\protect\hyperlink{Thm:#1}{Theorems{\kern 3pt}#1--#2}}
\newcommand{\tr}[0]{\text{tr}}
\newcommand{\shiftmath}[2]{\textnormal{\raisebox{#1}[#1][#1]{$#2$}}}

\newcommand{\hsp}[1]{{\kern #1pt}}
\newcommand{\rhoX}{\rho_\text{X}}
\newcommand{\cpre}{\overline{\hsp{-0.8}c\rule{0pt}{4.75pt}}}
\begin{document}
\title{X States of the Same Spectrum and Entanglement as All Two-Qubit States}
\author{Samuel R. Hedemann}
\noaffiliation
\date{\today}
\begin{abstract}
We present an explicit family of two-qubit X states with entanglement-preserving unitary (EPU) equivalence to the set of general states; that is, for any spectrum-entanglement combination achievable by general states, this family contains an X state of the same spectrum and entanglement. This idea was originally conjectured by the author and supported with strong numerical evidence in \href{http://arxiv.org/abs/1310.7038}{arXiv:1310.7038}. Then, in \href{http://dx.doi.org/10.1016/j.aop.2014.08.017}{Ann.~Phys.~351~(2014)~79}, the authors proved the existence of such two-qubit unitary transformations, but found the parameters to be transcendental, eluding explicit solution. Here, by a different method, we prove the existence of such transformations, obtain a compact implicit solution for them, and provide an exact, explicit form of the desired X-state family.
\end{abstract}
\maketitle
\section{\label{sec:I}Introduction}
Quantum entanglement \cite[]{Schr,EPR} is a special kind of nonlocal correlation that is much stronger than the typical correlations observed in everyday human experience. We have come to see it as a powerful resource for achieving many tasks better than what is classically possible, such as in quantum encryption \cite[]{BB84,B92,Eker}, quantum teleportation \cite[]{BBCJ,BPM1,BPM2}, and certain quantum-computing algorithms \cite[]{Feyn,Divi,Sho1,Sho2,Grov,Deu1,DeJo,CEMM}.

However, many of the benefits of entanglement rely on our ability to reduce sources of noise, and one strategy for this is to reduce the complexity of a given system.  To this end, others have shown benefits from preparing systems in certain special states, for which many density-matrix elements are zero \cite[]{YE04,AlJa,Wein,PABJ}. Furthermore, such states are easier to handle mathematically as well.

One important special state family is the X states, for which all potentially nonzero density-matrix elements are located on the main diagonal and antidiagonal as
\begin{Equation}                      {1}
\rho  = \left( {\begin{array}{*{20}c}
   {\rho _{1,1} } & 0 & 0 & {\rho _{1,4} }  \\
   0 & {\rho _{2,2} } & {\rho _{2,3} } & 0  \\
   0 & {\rho _{3,2} } & {\rho _{3,3} } & 0  \\
   {\rho _{4,1} } & 0 & 0 & {\rho _{4,4} }  \\
\end{array}} \right)\!,
\end{Equation}
where $\rho _{a,b}\equiv\langle a|\rho|b\rangle$ in a generic computational basis $\{|1\rangle,\ldots,|4\rangle\}$, and ket labels in our convention start on $1$ and do not imply the Fock basis \cite[]{Dira}. The density operator is $\rho =\sum_{j}p_{j}\rho_{j}$, with probabilities $p_{j}\in[0,1]$ such that $\sum_{j}p_{\shiftmath{0.5pt}{j}} =1$, and pure states $\rho_{j}\equiv|\psi_{j}\rangle\langle\psi_{j}|$ such that \smash{$\tr(\rho_{j})\equiv\sum\nolimits_{k=1}^{n}(\rho_{j})_{k,k}=1$},\hsp{-0.5} where\hsp{-0.5} $n\hsp{-0.5}=\hsp{-0.5}4$\hsp{-0.5} for\hsp{-0.5} two\hsp{-0.5} qubits.

The most important property of two-qubit X states regarding entanglement is that they are related to \textit{all} pure and mixed states by an entanglement-preserving unitary (EPU) transformation such that the transformed state has the same entanglement as the input state, a property called \textit{EPU equivalence}, as conjectured and supported with strong numerical evidence in \cite[]{HedX}, and later proven by \cite[]{MeMG}. For systems \textit{larger} than two qubits, \cite[]{HedX} proposed the \textit{true-generalized X (TGX) states}, which are actually \textit{non}X-shaped in general, and are so-named for their ability to generalize the EPU-equivalence property of two-qubit X states to larger systems.

However, while the conjecture of EPU equivalence from \cite[]{HedX} was proven true by \cite[]{MeMG}, that proof did not yield an explicit form for the EPU operator \smash{$U_{\text{EPU}_{\text{X}}}$} despite proving its existence, where we define general (and not necessarily local) EPU transformations as
\begin{Equation}                      {2}
U_{\text{EPU}}  \equiv U,\;\;\text{s.t.}\;\;\left\{\! {\begin{array}{*{20}l}
   {U^{\dag}  } &\!\! { = U^{ - 1} }  \\
   {E(U\rho U^{\dag}  )} &\!\! { = E(\rho ), }  \\
\end{array}} \right.
\end{Equation}
for valid entanglement measure $E(\rho)$. Thus, \smash{$U_{\text{EPU}}\rho U_{\text{EPU}}^{\dag}$} has the same entanglement and spectrum as $\rho$.

The main result of this paper is an explicit form for an X-state family that is EPU-equivalent to all general states, thus proving the conjecture of \cite[]{HedX} by an alternative method, and finally providing an explicit solution to achieve this transformation.

To quantify the entanglement of all two-qubit mixed states, we will use the \textit{concurrence} \cite[]{HiWo,Woot},
\begin{Equation}                      {3}
C(\rho ) \equiv \max \{0,\xi _1  - \xi _2  - \xi _3  - \xi _4 \},
\end{Equation}
where $\xi_{1}\geqslant\cdots\geqslant\xi_{4}$ are the eigenvalues of the Hermitian matrix $\sqrt {\sqrt \rho  \widetilde {\rho} \sqrt \rho  } $ (or square roots of the eigenvalues of nonHermitian matrix $\rho\widetilde {\rho}$), where $\widetilde{\rho}  \equiv (\sigma_2  \otimes \sigma _2 )\rho ^* (\sigma _2  \otimes \sigma _2 )$ with \smash{$\sigma_2 \equiv\binom{0\hsp{3}-i}{\hsp{-1.2}i\hsp{8}0}$}. If $\rho$ is an X state, \Eq{3} simplifies to
\begin{Equation}                      {4}
C(\rho ) = 2\max \{ 0,|\rho _{3,2} | - \sqrt {\rho _{4,4} \rho _{1,1} } ,|\rho _{4,1} | - \sqrt {\rho _{3,3} \rho _{2,2} } \,\},
\end{Equation}
valid for \textit{all} X states, both mixed and pure \cite[]{YuEb,WBPS}.
\section{\label{sec:II}X States of Exact EPU Equivalence to All States}
Given a general two-qubit state $\rho$ with eigenvalues $ \lambda_{1}\geqslant\cdots\geqslant\lambda_{4}$ and concurrence $C\equiv C(\rho)$ obtained from \Eq{3}, an X state of the same spectrum and concurrence is
\begin{Equation}                      {5}
\rhoX\equiv\frac{1}{2}\hsp{-2}\left(\hsp{-2} {\begin{array}{*{20}c}
   {\lambda _1  \hsp{-1}+\hsp{-1} \lambda _3  \hsp{-1}+\hsp{-1} \sqrt {\Omega\rule{0pt}{8.2pt}} } &\hsp{-3}  \cdot  &\hsp{-3}  \cdot  &\hsp{-3} {\sqrt {(\lambda _1  \hsp{-1}-\hsp{-1} \lambda _3 )^2  \hsp{-1}-\hsp{-1} \Omega\rule{0pt}{8.2pt}} }  \\
    \cdot  &\hsp{-3} {2\lambda _2 } &\hsp{-3}  \cdot  &\hsp{-3}  \cdot   \\
    \cdot  &\hsp{-3}  \cdot  &\hsp{-3} {2\lambda _4 } &\hsp{-3}  \cdot   \\
   {\sqrt {(\lambda _1 \hsp{-1}-\hsp{-1} \lambda _3 )^2  \hsp{-1}-\hsp{-1} \Omega\rule{0pt}{8.2pt}} } &\hsp{-3}  \cdot  &\hsp{-3}  \cdot  &\hsp{-3} {\lambda _1  \hsp{-1}+\hsp{-1} \lambda _3  \hsp{-1}-\hsp{-1} \sqrt {\Omega\rule{0pt}{8.2pt}} }  \\
\end{array}} \right)\hsp{-1},
\end{Equation}
where dots represent zeros and $\Omega \equiv \max \{ 0,Q\}$, where
\begin{Equation}                      {6}
Q \equiv (\lambda _1  - \lambda _3 )^2  - (C + 2\sqrt {\lambda _2 \lambda _4 } )^2 .
\end{Equation}

As given in \cite[]{HedX}, an exact EPU converting $\rho$ to $\rhoX$ is
\begin{Equation}                      {7}
U_{\text{EPU}_{\text{X}}}=\epsilon_{\rho_\text{X}}\epsilon_{\rho}^{\dag},
\end{Equation}
where $\epsilon_A$ is a unitary eigenvector matrix of $A$ such that \smash{$\epsilon_A^{\dag} A\epsilon_A=\Lambda$} is the diagonal matrix of eigenvalues $ a_{1}\geqslant\cdots\geqslant a_{4}$ of $A$ (the columns of $\epsilon_A$ are eigenvectors of $A$). Use singular value decompositions of $\rhoX$ and $\rho$ to ensure that \smash{$\epsilon_{\rhoX}$} and \smash{$\epsilon_\rho$} are unitary so that \smash{$U_{\text{EPU}_{\text{X}}}$} is also unitary.

\begin{figure}[H]
\centering
\vspace{-5pt}
\includegraphics[width=1.00\linewidth]{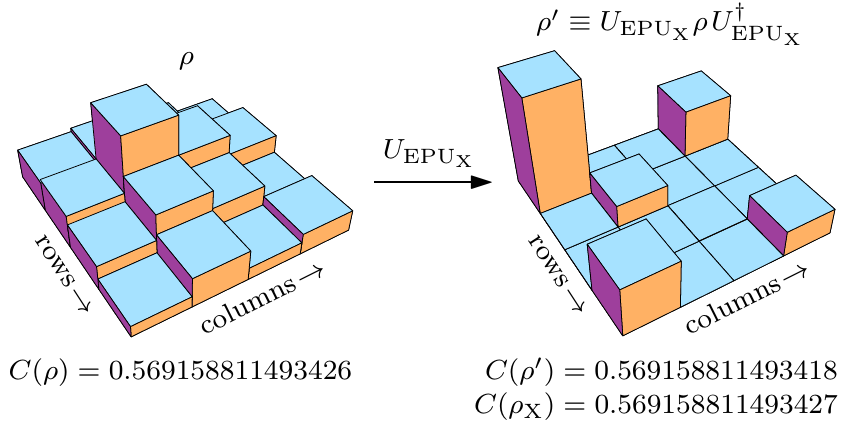}%
\vspace{-3pt}
\caption[]{(color online) Example of an arbitrary mixed two-qubit state $\rho$ transformed by\hsp{0.5} \smash{$U_{\text{EPU}_{\text{X}}}$}\hsp{-0.5} from \Eq{7} to get the EPU-equivalent\hsp{-1.75} X\hsp{-1.75} state\hsp{-1.75} \smash{$\rho'\hsp{-1}\equiv\hsp{-1} U_{\text{EPU}_{\text{X}}}\rho\hsp{1} U_{\text{EPU}_{\text{X}}}^{\dag}$} (phases not shown). Technically, $\rho'\hsp{-2}=\hsp{-2}\rhoX$ exactly, where $\rhoX$ is from \Eq{5}, but in numerical implementations, \Eq{5} has a slightly more accurate match of the spectrum and concurrence $C$ of $\rho$ since it involves fewer calculations. Typically, $\rhoX$ matches $C$ to $10$ digits of precision, while $\rho'$ matches $C$ to $9$ digits. For \textit{proof} that $\rhoX$ of \Eq{5} has \textit{exactly} the same $C$ and spectrum as $\rho$, see \Sec{III}.}
\label{fig:1}
\end{figure}

Typically, $\epsilon_{\rhoX}$ and $\epsilon_{\rho}$ will be numerically evaluated, even though they \textit{could} be found analytically since $n=4$. Thus, although this proves that $U_{\text{EPU}_{\text{X}}}$ can be found explicitly from \Eq{7} via explicit spectral decompositions of $\rho$ and $\rhoX$, it will be easier just to get it numerically, as seen in \Fig{1}. In practice, \Eq{5} gives us the explicit answer we seek without the need to use $U_{\text{EPU}_{\text{X}}}$ at all. Thus, given the spectrum and $C$ of any $\rho$, \textit{\Eq{5} is the exact result} of the desired EPU X transformation. 
\vspace{-1pt}
\vspace{-19.5pt}
\section{\label{sec:III}Proof of the Main Results}
\vspace{-5.5pt}
Here, it is useful to start by expanding \Eq{5} as
\begin{Equation}                      {8}
\rhoX  \hsp{-2}\equiv\hsp{-3} \left\{\hsp{-3} {\begin{array}{*{20}l}
   {\frac{1}{2}\hsp{-3}\left(\hsp{-2} {\begin{array}{*{20}c}
   {\lambda _1  \hsp{-2}+\hsp{-2} \lambda _3  \hsp{-2}+\hsp{-2} \sqrt{Q\rule{0pt}{8pt}}} &\hsp{-2}  \cdot  &\hsp{-2}  \cdot  &\hsp{-2} {C \hsp{-2}+\hsp{-2} 2\sqrt {\lambda _2 \lambda _4 \rule{0pt}{8pt}} }  \\
    \cdot  &\hsp{-2} {2\lambda _2 } &\hsp{-2}  \cdot  &\hsp{-2}  \cdot   \\
    \cdot  &\hsp{-2}  \cdot  &\hsp{-2} {2\lambda _4 } &\hsp{-2}  \cdot   \\
   {C \hsp{-2}+\hsp{-2} 2\sqrt {\lambda _2 \lambda _4 \rule{0pt}{8pt}} } &\hsp{-2}  \cdot  &\hsp{-2}  \cdot  &\hsp{-2} {\lambda _1  \hsp{-2}+\hsp{-2} \lambda _3  \hsp{-2}-\hsp{-2} \sqrt{Q\rule{0pt}{8pt}}}  \\
\end{array}}\hsp{-1} \right)\!\hsp{-1};} &\hsp{-3} {Q \hsp{-2}\geqslant\hsp{-2} 0}  \\
   {\frac{1}{2}\hsp{-3}\left(\hsp{-2} {\begin{array}{*{20}c}
   {\lambda _1  \hsp{-2}+\hsp{-2} \lambda _3  } &\hsp{-2}  \cdot  &\hsp{-2}  \cdot  &\hsp{-2} {\lambda _1  \hsp{-2}-\hsp{-2} \lambda _3}  \\
    \cdot  &\hsp{-2} {2\lambda _2 } &\hsp{-2}  \cdot  &\hsp{-2}  \cdot   \\
    \cdot  &\hsp{-2}  \cdot  &\hsp{-2} {2\lambda _4 } &\hsp{-2}  \cdot   \\
   {\lambda _1  \hsp{-2}-\hsp{-2} \lambda _3} &\hsp{-2}  \cdot  &\hsp{-2}  \cdot  &\hsp{-2} {\lambda _1  \hsp{-2}+\hsp{-2} \lambda _3  }  \\
\end{array}}\hsp{-1} \right)\!\hsp{-1};} &\hsp{-3} {Q \hsp{-2}<\hsp{-2} 0,}  \\
\end{array}} \right.\hsp{-4}
\end{Equation}
where we used the definition of $Q$ in \Eq{6} to alter the off-diagonals in the $Q\geqslant 0$ case.  As we will prove soon, $C=0$ in the $Q<0$ case, so we \textit{could} have used the \textit{diagonal} state $\text{diag}\{ \lambda _1 ,\lambda _2 ,\lambda _4 ,\lambda _3 \}$ to achieve the same goal with even more simplicity; however the form in \Eq{8} allows the continuity between the $Q$ cases that permits their unification to the compact form in \Eq{5}.

To \textit{prove} that $\rhoX$ of \Eq{8} is a universal family of X states for entanglement, \textit{we must show that} $\rhoX$ \textit{contains all combinations of spectrum, concurrence, and $Q$ value ($\Lambda CQ$) that exist for the set of general two-qubit quantum states}. 

The spectrum matching will permit unitary equivalence, which, combined with the concurrence preservation, will enable the EPU-equivalence property. Then, if we show that each $Q$ case in \Eq{8} permits all $\Lambda C$ combinations for that $Q$ value, that will complete the proof that $\rho_{\text{X}}$ is EPU-equivalent to all states $\rho$, also proving that \Eq{7} is the form of such an EPU transformation.

Here, we start with a top-down proof in \Sec{III.A}, simply verifying that $\rho_{\text{X}}$ satisfies the above properties. Then, \Sec{III.B} shows how $\rho_{\text{X}}$ was obtained in the first place.
\subsection{\label{sec:III.A}Proof that $\rho_{\text{X}}$ is EPU-Equivalent to All States}
To prove that this transformation works on all states, simply put \Eq{8} into \Eq{4}. If $\rho_\text{X}$ is EPU-equivalent to the set of all two-qubit states, then calculating its concurrence should yield $C$ for all possible spectrum-concurrence ($\Lambda C$) combinations for each $Q$ case, where $\lambda_{1}\geqslant\cdots\geqslant\lambda_{4}$.  We \textit{also} need to show that the $Q$ cases are exhaustive for all $\rho$ and that each $Q$ case covers all $\Lambda C$ combinations, but we will address that after treating spectrum and concurrence preservation.
\\
\vspace{-4pt}
\\
\textbf{Proof that Spectrum is Preserved:} In both cases of \Eq{8}, $\text{det}(\lambda I-\rho_{\text{X}})=(\lambda-\lambda_1)(\lambda-\lambda_2)(\lambda-\lambda_3)(\lambda-\lambda_4)=0$, proving that $\rho_{\text{X}}$ has the same spectrum as $\rho$, and that they are unitarily equivalent.
\\
\vspace{-4pt}
\\
\textbf{Proof\hsp{-3} that\hsp{-3} Concurrence\hsp{-3} is\hsp{-3} Preserved:}\hsp{-3} In the $Q\geqslant 0$ case, putting \Eq{8} into \Eq{4} gives
\begin{Equation}                      {9}
\begin{array}{*{20}l}
   {C(\rhoX )} &\!\! { = 2\max \{ 0, - \sqrt {{\rhoX }_{4,4} {\rhoX }_{1,1} } ,\frac{{C + 2\sqrt {\lambda _2 \lambda _4 } }}{2} - \sqrt {\lambda _2 \lambda _4 } \} }  \\
   {} &\!\! { = 2\max \{ 0,\frac{C}{2}\} }  \\
   {} &\!\! { = C,}  \\
\end{array}
\end{Equation}
which shows that $\rhoX$ is indeed EPU-equivalent to $\rho$ when $Q\geqslant 0$, since $C$ was computed from $\rho$ to get \Eq{8}, and because \Eq{8} allows all combinations of eigenvalues (in descending order, which does not reduce the generality). 

In the $Q<0$ case, $\rho$ is always separable (which we prove soon), so $C=0$, and putting \Eq{8} into \Eq{4} gives
\begin{Equation}                      {10}
\begin{array}{*{20}l}
   {C(\rhoX )} &\!\! { = \max\{0,\lambda _1  - \lambda _3 -2\sqrt {\lambda _2 \lambda _4 }\};\;Q<0 }  \\
   {} &\!\! { = 0}  \\
   {} &\!\! { = C,}  \\
\end{array}
\vspace{-2pt}
\end{Equation}
where we used the fact that \smash{$\lambda _1  - \lambda _3 -2\sqrt {\lambda _2 \lambda _4 }<0$} when $Q<0$, which we will also prove soon.

Thus, both $Q$ cases of $\rhoX$ preserve spectrum and concurrence of all $\rho$ that have those $Q$ values (where we still need to prove \Eq{10} and show that the $Q$ cases are exhaustive and that $C=0$ for all states when $Q<0$).
\\
\vspace{-4pt}
\\
\textbf{Proof\hsp{-1.0} that\hsp{-1.0} the\hsp{-1.0} $Q$\hsp{-1.0} Cases\hsp{-1.0} Are\hsp{-1.0} $\Lambda C$-Exhaustive:}\hsp{-0.5} Here, we need to show that (i) all states $\rho$ only fall into the two $Q$ cases in \Eq{8}, and that (ii) both of those forms of $\rhoX$ admit all spectrum-concurrence combinations for those $Q$ cases (which requires that we prove $C\hsp{-1}=\hsp{-1}0$ $\forall\rho$ when $Q<0$).  For these proofs, we use the notation that $\Lambda\!\equiv\!\text{diag}\{\lambda_k\}$ where $\{\lambda_k\}\!\equiv\!\{\lambda_k\}|_{k=1}^{k=4}\!\equiv\!\{\lambda_1,\lambda_2,\lambda_3,\lambda_4\}$.
\begin{itemize}[leftmargin=*,labelindent=8pt]\setlength\itemsep{0pt}
\item[\textbf{i.}]\hypertarget{LambdaCExhaustive:i}{}\textbf{Proof~that~All~$\rho$~Qualify~as~One~of~the~$Q$ Cases:} Given the dependence of $Q$ as
\begin{Equation}                      {11}
\begin{array}{*{20}l}
   Q &\!\! { \equiv Q(\lambda _1 ,\lambda _2 ,\lambda _3 ,\lambda _4 ,C)}  \\
   {} &\!\!  \equiv (\lambda _1  - \lambda _3 )^2  - (C + 2\sqrt {\lambda _2 \lambda _4 } )^2 ,  \\
\end{array}
\end{Equation}
and the fact that
\begin{Equation}                      {12}
\left\{ {\begin{array}{*{20}l}
   {\text{$Q$ Case 1:}} & {Q \geqslant 0}  \\
   {\text{$Q$ Case 2:}} & {Q < 0,}  \\
\end{array}} \right.
\end{Equation}
then we can make the following conclusions:
    \begin{itemize}[leftmargin=*,labelindent=4pt]\setlength\itemsep{0pt}
    \item[\textbf{a.}]\hypertarget{LambdaCExhaustive:i.a}{}The fact that $Q$ is \textit{real} for every combination of $\{\lambda_k\}$ and $C$, meaning $Q^{*}=Q\;\forall\,C,\{\lambda_k\}$, proves that the two mutually exclusive $Q$ cases are exhaustive values of $Q$ since the two cases cover all possibilities for $Q$.
    \item[\textbf{b.}]\hypertarget{LambdaCExhaustive:i.b}{}Since every $\rho$ has a spectrum $\{\lambda_k\}$ and concurrence value $C$ (even if zero), then for every $\rho$, there exists a value of $Q$, meaning $\exists\, Q\;\forall\,\rho$.
    \item[\textbf{c.}]\hypertarget{LambdaCExhaustive:i.c}{}Therefore, due to \hyperlink{LambdaCExhaustive:i.a}{Conclusion a} and \hyperlink{LambdaCExhaustive:i.b}{Conclusion b}, every $\rho$ falls into exactly one of the two $Q$ cases.
    \end{itemize}
Here, we have only shown that $Q$ is not limited by its input arguments and that all $\rho$ fall into one of the two $Q$ cases, but now we need to show that the cases of $Q$ in \Eq{12} do not limit the \textit{$\Lambda C$ combinations}. Specifically, \textit{does the form of} $\rhoX$ \textit{for each $Q$ case in \Eq{8} allow all possible $C$ values for a given spectrum?}
\item[\textbf{ii.}]\hypertarget{LambdaCExhaustive:ii}{}\textbf{Proof that Each $Q$-Case $\rhoX$ Admits All $\Lambda C$ Combinations:~~}
    \begin{itemize}[leftmargin=*,labelindent=4pt]\setlength\itemsep{0pt}
    \item[\textbf{a.}]\hypertarget{LambdaCExhaustive:ii.a}{}For $Q\geqslant 0$, the result in \Eq{9} that $C(\rhoX)=C$ $\forall\,\{\lambda_k\}$ proves that this $Q$-case $\rhoX$ admits all $\Lambda C$ combinations.
    \item[\textbf{b.}]\hypertarget{LambdaCExhaustive:ii.b}{}For $Q<0$, consider the following facts:
        \begin{itemize}[leftmargin=*,labelindent=4pt]\setlength\itemsep{0pt}
        \item[\textbf{1.}]\hypertarget{LambdaCExhaustive:ii.b.1}{}At the edge of the $Q\geqslant 0$ case, if $Q=0$, then according to \Eq{6}, $C + 2\sqrt {\lambda _2 \lambda _4 }=\lambda _1  - \lambda _3$, so then $\rhoX$ becomes (still in the $Q\geqslant 0$ case) 
\begin{Equation}                      {13}
\hsp{35}\begin{array}{*{20}l}
   {\rhoX=} &\!\! {\frac{1}{2}\hsp{-3}\left(\hsp{-2} {\begin{array}{*{20}c}
   {\lambda _1  \hsp{-2}+\hsp{-2} \lambda _3  } &\hsp{-2}  \cdot  &\hsp{-2}  \cdot  &\hsp{-2} {\lambda _1  \hsp{-2}-\hsp{-2} \lambda _3}  \\
    \cdot  &\hsp{-2} {2\lambda _2 } &\hsp{-2}  \cdot  &\hsp{-2}  \cdot   \\
    \cdot  &\hsp{-2}  \cdot  &\hsp{-2} {2\lambda _4 } &\hsp{-2}  \cdot   \\
   {\lambda _1  \hsp{-2}-\hsp{-2} \lambda _3} &\hsp{-2}  \cdot  &\hsp{-2}  \cdot  &\hsp{-2} {\lambda _1  \hsp{-2}+\hsp{-2} \lambda _3  }  \\
\end{array}}\hsp{-1} \right)\!,}  \\
\end{array}
\end{Equation}
which is a \textit{maximally entangled mixed state} (MEMS) \cite[]{IsHi,ZiBu,HoBM} with respect to (wrt) a given spectrum, as proven in \cite[]{VeAM,WNGK} up to a local-unitary permutation by $\binom{0\hsp{3}1}{1\hsp{3}0}\otimes\binom{1\hsp{3}0}{0\hsp{3}1}$ on \Eq{13}, and actually has this same form for all $Q$.
        \item[\textbf{2.}]\hypertarget{LambdaCExhaustive:ii.b.2}{}Putting \Eq{13} into \Eq{4} gives the exact concurrence of any MEMS wrt spectrum as
\begin{Equation}                      {14}
\hsp{35}C_{\text{MEMS}_{\Lambda}} \equiv\max\{0,\lambda _1  - \lambda _3 -2\sqrt {\lambda _2 \lambda _4 }\},
\end{Equation}
        (also valid for any $Q$), which means that
\begin{Equation}                      {15}
\hsp{35}\cpre\equiv\lambda _1  - \lambda _3 -2\sqrt {\lambda _2 \lambda _4 }
\end{Equation}
is the \textit{minimum average preconcurrence} \cite[]{Woot} of a MEMS wrt spectrum, and can be negative (for example, when all eigenvalues are $\frac{1}{4}$).
        \item[\textbf{3.}]\hypertarget{LambdaCExhaustive:ii.b.3}{}No state can have a larger $C$ than $C_{\text{MEMS}_{\Lambda}}$;
\begin{Equation}                      {16}
\hsp{35}0\leqslant C(\rho)\leqslant C_{\text{MEMS}_{\Lambda}}\;\;\forall\,\rho, \;\forall Q.
\end{Equation}
        \item[\textbf{4.}]\hypertarget{LambdaCExhaustive:ii.b.4}{}Focusing now on the $Q<0$ case, solve for any conditions that $Q<0$ implies for $C$:
\begin{Equation}                      {17}
\hsp{35}\begin{array}{*{20}l}
   {\hsp{109.5}Q} &\!\!\!  <  &\!\!\! 0  \\
   {(\lambda _1  \hsp{-1}-\hsp{-1} \lambda _3 )^2  \hsp{-1}-\hsp{-1} (C \hsp{-1}+\hsp{-1} 2\sqrt {\lambda _2 \lambda _4 } )^2} &\!\!\!  <  &\!\!\! 0  \\
   {\hsp{65}C \hsp{-1}+\hsp{-1} 2\sqrt {\lambda _2 \lambda _4 } } &\!\!\!  >  &\!\!\! {\lambda _1  \hsp{-1}-\hsp{-1} \lambda _3 }  \\
   {\hsp{109.5}C} &\!\!\!  >  &\!\!\! {\lambda _1  \hsp{-1}-\hsp{-1} \lambda _3  \hsp{-1}-\hsp{-1} 2\sqrt {\lambda _2 \lambda _4 } }  \\
   {\hsp{109.5}C} &\!\!\!  >  &\!\!\! {\cpre,}  \\
\end{array}
\end{Equation}
        which means that
\begin{Equation}                      {18}
\hsp{40}(Q < 0) \Rightarrow (C > \cpre) \Rightarrow \left\{ {\begin{array}{*{20}l}
   {C > C_{\text{MEMS}_{\Lambda}} ;} &\!\! {\cpre \geqslant 0}  \\
   {C > \cpre;} &\!\! {\cpre < 0,}  \\
\end{array}} \right.
\end{Equation}
since $\cpre = C_{\text{MEMS}_{\Lambda}}$ when $\cpre \geqslant 0$ by \Eqs{14}{15}.
        \item[\textbf{5.}]\hypertarget{LambdaCExhaustive:ii.b.5}{}The $\cpre \geqslant 0$ case of \Eq{18} means that $Q<0$ implies that $C>C_{\text{MEMS}_{\Lambda}}$, \textit{which is never possible} because $\max(C)=C_{\text{MEMS}_{\Lambda}}$ from \Eq{16}.  Therefore, the only case of \Eq{18} that can apply to physical states is the $\cpre < 0$ case, which yields 
\begin{Equation}                      {19}
\hsp{40}(Q < 0) \Rightarrow (C > \cpre) \Rightarrow (\cpre < 0) \;\;\forall\,\rho.
\end{Equation}
        \item[\textbf{6.}]\hypertarget{LambdaCExhaustive:ii.b.6}{}From \Eqs{14}{15}, we know that 
\begin{Equation}                      {20}
\hsp{35}(\cpre < 0)\Rightarrow (C_{\text{MEMS}_{\Lambda}}=0).
\end{Equation}
        \item[\textbf{7.}]\hypertarget{LambdaCExhaustive:ii.b.7}{}Then, putting $C_{\text{MEMS}_{\Lambda}}=0$ from \Eq{20} into \Eq{16},
\begin{Equation}                      {21}
\hsp{35}0\leqslant C(\rho)\leqslant 0\;\;\forall\,\rho,\;\;\text{when}\;\; \cpre < 0.
\end{Equation}
        \item[\textbf{8.}]\hypertarget{LambdaCExhaustive:ii.b.8}{}By \Eq{19} and \Eq{21}, we obtain
\begin{Equation}                      {22}
\hsp{35}(Q<0)\Rightarrow (C(\rho)=0)\;\;\forall\,\rho.
\end{Equation}
        \item[\textbf{9.}]\hypertarget{LambdaCExhaustive:ii.b.9}{}From the fact in \Eq{19} that $Q<0$ implies that $\cpre\hsp{-0.75}<\hsp{-0.75}0$, then by \Eq{15}, $\lambda _1  - \lambda _3 -2\sqrt {\lambda _2 \lambda _4 }\hsp{-0.75}<\hsp{-0.75}0$, which \textit{proves} the claim in \Eq{10} that $C(\rhoX)=0$ for all spectra $\Lambda$ for which $Q<0$. Thus, comparing the now-proven result in \Eq{10} with the result of \Eq{22} regarding \textit{general} states $\rho$ proves that: 
\begin{Equation}                      {23}
\hsp{40}\begin{array}{*{20}l}
   {\parbox{2.0in}{The $Q<0$ case of $\rhoX$ in \Eq{8} \textit{does} exhaust all possible $\Lambda C$ combinations when $Q<0$.}}  \\
\end{array}
\end{Equation}
        \end{itemize}
    \end{itemize}
Therefore, since \hyperlink{LambdaCExhaustive:ii.a}{(a)} and \hyperlink{LambdaCExhaustive:ii.b}{(b)} prove that $\rhoX$ admits all possible spectrum and concurrence combinations for all possible $Q$ cases, the claim of \hyperlink{LambdaCExhaustive:ii}{(ii)} is proven true.
\end{itemize}
Finally, since Claims \hyperlink{LambdaCExhaustive:i}{(i)} and \hyperlink{LambdaCExhaustive:ii}{(ii)} have both been proven true, this completes the proof that the state family $\rhoX$ in \Eq{8} (and thus in \Eq{5}) is fully EPU-equivalent to the set of all states $\rho$, since the set of all possible $\Lambda CQ$ combinations achievable by $\rho$ is also achievable by $\rhoX$. 

Note that throughout this paper, by ``all $C$ values,'' we mean \textit{physical} $C$, meaning that we only consider $C\in[0,\max\{0,\lambda _1  - \lambda _3 -2\sqrt {\lambda _2 \lambda _4 }\}]$, as indicated in \Eq{16}.
\subsection{\label{sec:III.B}Motivation of $\rho_{\text{X}}$ as an Ansatz Prior to Proof}
Now that we have \textit{proven} the EPU equivalence of $\rhoX$ of \Eq{5} to general states $\rho$, our task here is to show how that state family was obtained in the first place.
\subsubsection{\label{sec:III.B.1}Finding an EPU-Compatible Spectral X Decomposition}
First, note that X states $\rhoX$ of a given spectrum $\Lambda\equiv\text{diag}\{\lambda_{1},\ldots,\lambda_{4}\}$ where $\lambda_{1}\geqslant\cdots\geqslant\lambda_{4}$ can be made as
\begin{Equation}                      {24}
\rhoX=\epsilon _{\rhoX } \Lambda \epsilon _{\rhoX}^\dag,
\end{Equation}
where $\epsilon _{\rhoX }$ is a unitary eigenvector matrix whose columns are X states.  Since any relative phase of $\rhoX$ can be absorbed into the total EPU transformation, we only need to consider real-valued $\rhoX$ and $\epsilon _{\rhoX }$. Therefore, the general set of eigenstates we need are X-form $\theta$ states \cite[]{HedE},
\begin{Equation}                      {25}
\left(\hsp{-0} {\begin{array}{*{20}c}
   {c_\alpha  }  \\
    \cdot   \\
    \cdot   \\
   {s_\alpha  }  \\
\end{array}}\hsp{-0} \right)\hsp{-3.0},
\left(\hsp{-3} {\begin{array}{*{20}c}
   {s_\alpha  }  \\
    \cdot   \\
    \cdot   \\
   { - c_\alpha  }  \\
\end{array}}\hsp{-3} \right)\hsp{-3.0},
\left(\hsp{-0} {\begin{array}{*{20}c}
    \cdot   \\
   {c_\beta  }  \\
   {s_\beta  }  \\
    \cdot   \\
\end{array}}\hsp{-0} \right)\hsp{-3.0},
\left(\hsp{-3} {\begin{array}{*{20}c}
    \cdot   \\
   {s_\beta  }  \\
   { - c_\beta  }  \\
    \cdot   \\
\end{array}}\hsp{-3} \right)\hsp{-3.0},
\end{Equation}
where $c_{\theta}\equiv\cos(\theta)$, $s_{\theta}\equiv\sin(\theta)$, and $\alpha,\beta\in[0,\frac{\pi}{2}]$. Each state in \Eq{25} can assume any entanglement, but note that it consists of two pairs of related states, due to the simultaneous conditions of X form and unitarity.

Since $\Lambda$ already fully specifies the spectrum in \Eq{24}, and the X form of the states in \Eq{25} already enforces the X form of $\rhoX$, the parameters $\alpha$ and $\beta$ are the freedom with which we will ensure entanglement preservation to achieve $C(\rhoX)=C(\rho)$.  However, we must ask, \textit{is there a preferred way to assign the eigenvectors of \Eq{25} to the eigenvalues?}  In fact, there \textit{is} a preferred choice, since some choices can cause unacceptable restrictions on the spectrum-concurrence combinations.

The key to finding an X state capable of all $\Lambda C$ combinations lies in comparing the MEMS of \Eq{13} (for all $Q$ here) to the eigenvectors of \Eq{25}. The positions and coefficients of eigenvalues in \Eq{13} yield the desired order as
\begin{Equation}                      {26}
|\epsilon_{1}  \rangle  \hsp{-2}\equiv\hsp{-5} \left(\hsp{-0} {\begin{array}{*{20}c}
   {c_\alpha  }  \\
    \cdot   \\
    \cdot   \\
   {s_\alpha  }  \\
\end{array}}\hsp{-0} \right)\hsp{-3.5},
\hsp{1.50}|\epsilon_{2}  \rangle  \hsp{-2}\equiv\hsp{-5} \left(\hsp{-0} {\begin{array}{*{20}c}
    \cdot   \\
   {c_\beta  }  \\
   {s_\beta  }  \\
    \cdot   \\
\end{array}}\hsp{-0} \right)\hsp{-3.5},
\hsp{1.50}|\epsilon_{3}  \rangle  \hsp{-2}\equiv\hsp{-5} \left(\hsp{-3} {\begin{array}{*{20}c}
   {s_\alpha  }  \\
    \cdot   \\
    \cdot   \\
   { - c_\alpha  }  \\
\end{array}}\hsp{-3} \right)\hsp{-3.5},
\hsp{1.50}|\epsilon_{4}  \rangle  \hsp{-2}\equiv\hsp{-5} \left(\hsp{-3} {\begin{array}{*{20}c}
    \cdot   \\
   {s_\beta  }  \\
   { - c_\beta  }  \\
    \cdot   \\
\end{array}}\hsp{-3} \right)\hsp{-3.5},
\end{Equation}
where the second and third states of \Eq{25} have been swapped, and ket labels have subscripts that match their respective eigenvalues. Thus, for the MEMS of \Eq{13}, setting $\alpha=\frac{\pi}{4}$ and $\beta=0$ in \Eq{26} would produce a proper set of eigenvectors for those states.

For the most general $\rhoX$, we simply leave $\alpha$ and $\beta$ free (at first), and define its eigenvector matrix as
\begin{Equation}                      {27}
\epsilon _{\rhoX}  \equiv \sum\limits_{k = 1}^4 {|\epsilon _k \rangle \langle k|} ,
\end{Equation}
which, when put into \Eq{24}, yields
\begin{Equation}                      {28}
\begin{array}{l}
 \rhoX  =  \\ 
 \hsp{-2}\left(\hsp{-3} {\begin{array}{*{20}c}
   {\lambda _1 c_\alpha ^2  \hsp{-2}+\hsp{-2.5} \lambda _3 s_\alpha ^2 } &\hsp{-7}  \cdot  &\hsp{-4}  \cdot  &\hsp{-7} {(\hsp{-0.5}\lambda _1  \hsp{-2}-\hsp{-2} \lambda _3 \hsp{-0.5})s_\alpha  c_\alpha  }  \\
    \cdot  &\hsp{-7} {\lambda _2 c_\beta ^2  \hsp{-2}+\hsp{-2.5} \lambda _4 s_\beta ^2 } &\hsp{-4} {(\hsp{-0.5}\lambda _2  \hsp{-2}-\hsp{-2} \lambda _4 \hsp{-0.5})s_\beta  c_\beta  } &\hsp{-7}  \cdot   \\
    \cdot  &\hsp{-7} {(\hsp{-0.5}\lambda _2  \hsp{-2}-\hsp{-2} \lambda _4 \hsp{-0.5})s_\beta  c_\beta  } &\hsp{-4} {\lambda _2 s_\beta ^2  \hsp{-2}+\hsp{-2.5} \lambda _4 c_\beta ^2 } &\hsp{-7}  \cdot   \\
   {(\hsp{-0.5}\lambda _1  \hsp{-2}-\hsp{-2} \lambda _3 \hsp{-0.5})s_\alpha  c_\alpha  } &\hsp{-7}  \cdot  &\hsp{-4}  \cdot  &\hsp{-7} {\lambda _1 s_\alpha ^2  \hsp{-2}+\hsp{-2.5} \lambda _3 c_\alpha ^2 }  \\
\end{array}}\hsp{-4} \right)\hsp{-2.5}. \\ 
 \end{array}\!
\end{Equation}

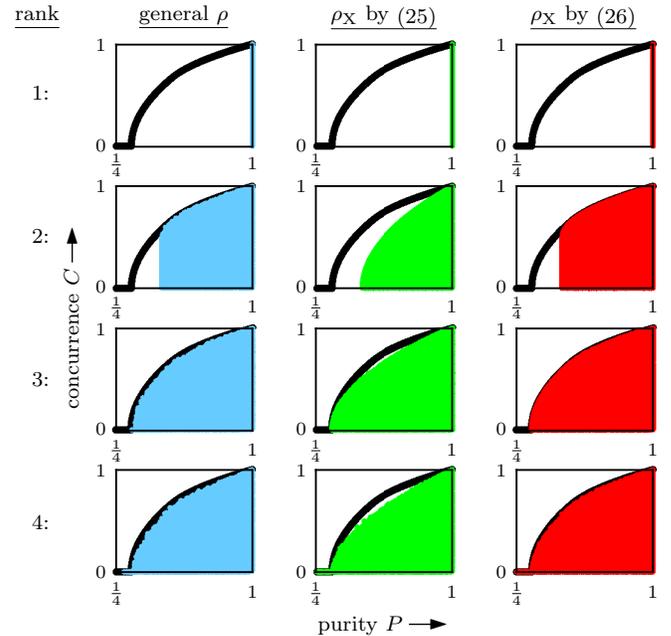
\begin{figure}[H]
\centering
\vspace{-1pt}
\includegraphics[width=1.00\linewidth]{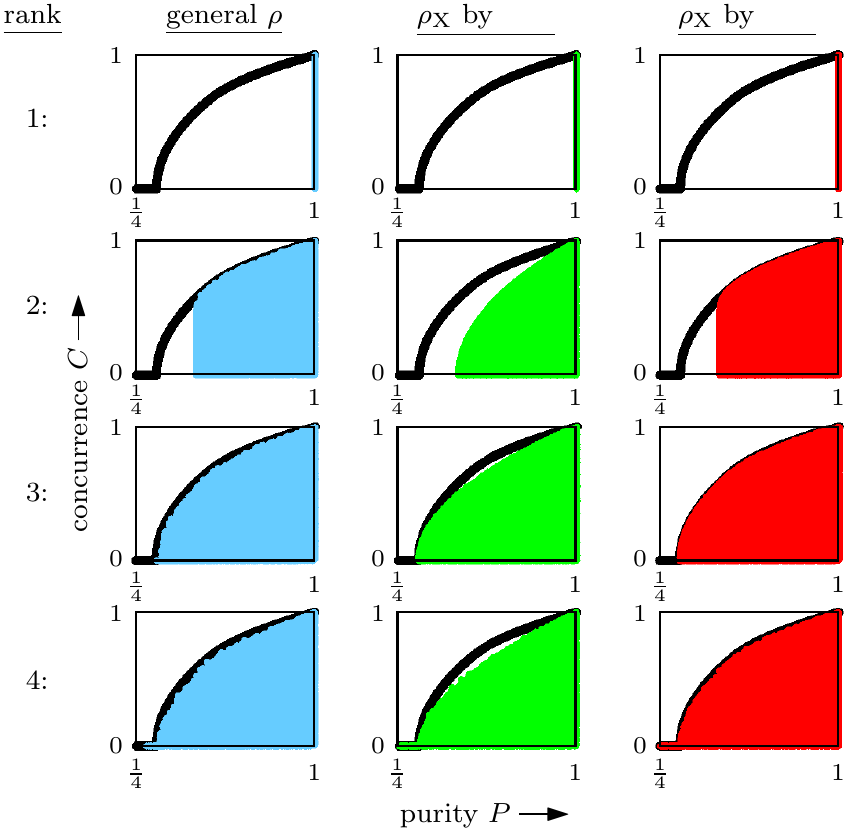}%
\vspace{-12pt}
\setlength{\unitlength}{0.01\linewidth}
\begin{picture}(100,0)
\put(49,94.0){\footnotesize \phantom{$\rhoX$ by }\Eq{25}}
\put(79.7,94.0){\footnotesize \phantom{$\rhoX$ by }\Eq{26}}
\end{picture}
\vspace{-12pt}
\caption[]{(color online) Plot of concurrence $C$ versus purity $P$ for $10^6$ arbitrary states for each rank for each state family; general $\rho$, X states with the eigenvector order of \Eq{25}, and X states with the eigenvector order of \Eq{26} (used in $\rhoX$ of \Eq{28}). The black curves are for the MEMS of \Eq{13} (for all $Q$ here), as in \cite[]{HedX}. The apparent inability of rank-2 (and maybe ranks 3 and 4) X states based on \Eq{25} to achieve the same $CP$ combinations as general $\rho$ may be evidence that X states based on \Eq{25} cannot achieve EPU equivalence with general $\rho$ since there are general rank-2 $\rho$ with $C$ too high for that family of X states to achieve. The $\rhoX$ of \Eq{28} based on \Eq{26} have no such limitation, so there is no reason to disqualify them based on this (limited) test. (This is merely an exploratory numerical motivation leading to an ansatz, but it paid off since the ansatz was proven correct as shown in \Sec{III.A}.)}
\label{fig:2}
\end{figure}

To motivate using \Eq{26} (and thus \Eq{28}) instead of \Eq{25}, \Fig{2} plots some necessary (but not sufficient) tests for EPU equivalence to general $\rho$ as numerical explorations.
\vspace{-3pt}
The \textit{most general} test of which eigenvector ordering to use would be to analytically check all possible \textit{combinations} of ordered eigenvalues and $C$ and see if any limitations occur when comparing their performance to that of general states $\rho$, and furthermore, to extend the candidate eigenvector sets beyond the two considered here. However, since explicit solution of $C$ for general states is extremely complicated, and since even approximating such a test numerically with a grid search would be almost intractable, we instead limit ourselves to approximations of the following necessary test.

As seen in \Fig{2}, we merely plot concurrence $C$ against purity $P\equiv\tr{(\rho^{2})}$ for many arbitrary states in each of three families; general states $\rho$, spectral X states with ordered eigenvectors as in \Eq{25}, and spectral X states with the eigenvectors of \Eq{26}. If any family does not cover the same area as $\rho$ under the $CP$ curve for any rank, then such a family may lack EPU equivalence to general $\rho$.  Again, a necessary and sufficient test for EPU equivalence would analytically check all combinations of $C$ with all nontrivial power sums simultaneously (\smash{$P_{m}\equiv\sum\nolimits_{\shiftmath{0.9pt}{k\!=\!1}}^{4}\lambda_{k}^{m}$} for $m=2,3,4$), not merely purity \smash{$P=P_{2}=\lambda_{1}^{2}+\lambda_{2}^{2}+\lambda_{3}^{2}+\lambda_{4}^{2}$}, but here we are merely approximating a necessary test to motivate an ansatz. This was an effective strategy because we were then able to \textit{prove} that ansatz to be true from a top-down approach.

Since the states made by \Eq{25} appear to have a deficiency for rank $2$ (and maybe ranks 3 and 4), they do not likely have EPU equivalence to general $\rho$. (Since the test is numerical this is not conclusive, but it is enough to motivate an ansatz.) Meanwhile, the $\rhoX$ of \Eq{28}, based on \Eq{26}, appears to have no limitations in covering the same $CP$ area as general states for each rank.  Again, we would need to (analytically) check $C$ for all power sums of eigenvalues simultaneously, not merely purity, but this test gave us a good reason to prefer \Eq{26} over \Eq{25}, and showed that there is at least no apparent reason not to suspect \Eq{26} of being the desired order.  Therefore, the $\rhoX$ of \Eq{28} is the family we chose as a general starting point. (Note that in \Fig{2}, states of rank $R$ have minimum purity \smash{$P_{\min}\hsp{-1}=\hsp{-1}\frac{1}{R}$} which accounts for vertical ``walls'' of data in the plots, and all states with purity less than \smash{$\frac{1}{n-1}\hsp{-1}=\hsp{-1}\frac{1}{3}$} are guaranteed to be separable \cite[]{ZHSL}, corresponding to the horizontal portions of data with \smash{$C_{\text{MEMS}_{\Lambda}}\hsp{-1}=\hsp{-1}0$}.)

At this stage, the states in \Eq{28} are still a bit too general to be useful, so we now seek values of $\alpha$ and $\beta$ to enable the EPU equivalence of $\rhoX$ to general $\rho$.
\subsubsection{\label{sec:III.B.2}Finding the Entanglement-Preserving Parameters}
Here, we want to investigate the effects of $\alpha$ and $\beta$ in \Eq{28} on its ability to achieve EPU equivalence to general states.  First, putting \Eq{28} into \Eq{4} gives
\begin{Equation}                      {29}
\begin{array}{*{20}l}
   {C(\alpha ,\beta ) \equiv } &\!\! {2\max \left\{ {\rule{0pt}{9pt}} \right.\hsp{-2}0,}  \\
   {} &\!\! {\frac{{\lambda _2  - \lambda _4 }}{2}s_{2\beta }  -\hsp{-1} \sqrt {(\lambda _1 s_\alpha ^2  + \lambda _3 c_\alpha ^2 )(\lambda _1 c_\alpha ^2  + \lambda _3 s_\alpha ^2 )\rule{0pt}{8.0pt}} ,}  \\
   {} &\!\! {\frac{{\lambda _1  - \lambda _3 }}{2}s_{2\alpha }  -\hsp{-1} \sqrt {(\lambda _2 s_{\smash{\beta}\vphantom{\alpha}} ^2  + \lambda _4 c_{\smash{\beta}\vphantom{\alpha}} ^2 )(\lambda _2 c_{\smash{\beta}\vphantom{\alpha}} ^2  + \lambda _4 s_{\smash{\beta}\vphantom{\alpha}} ^2 )\rule{0pt}{8.0pt}} \left. {\rule{0pt}{9pt}} \right\}\hsp{-1}.}  \\
\end{array}
\end{Equation}
To get a feel for \Eq{29}, we do numerical exploration as follows. From a given arbitrary input state $\rho$, we harvest its spectrum $\Lambda$ and concurrence $C$, then use $\Lambda$ to get $C(\alpha ,\beta )$ from \Eq{29} with a grid search over $\alpha$ and $\beta$, while using $C$ to plot the plane of the correct value.  The places where $C(\alpha ,\beta )$ intersects the $C$ plane show us what values of $\alpha$ and $\beta$ allow $\rhoX$ to preserve the entanglement of $\rho$.

\begin{figure}[H]
\centering
\includegraphics[width=1.00\linewidth]{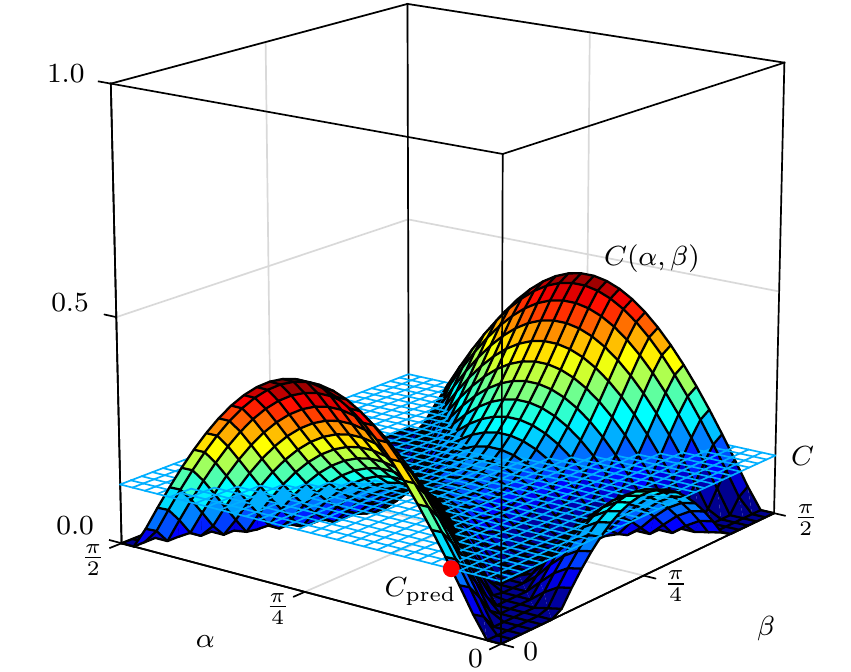}%
\vspace{-4pt}
\caption[]{(color online) Example of approximation of $C(\alpha,\beta)$ from \Eq{29} for a particular arbitrary rank-$4$ spectrum and target concurrence $C$ (which would be inputs to \Eq{5}). The value $C_{\text{pred}}$ predicted by using \Eq{30} and \Eq{32} in \Eq{29} is shown as the red dot, and the planar surface shows the target $C$. If the red dot always lies on an intersection of the plane $C$ and the surface $C(\alpha,\beta)$, then the predicted concurrence $C_{\text{pred}}$ is correct.  Repeating this test over $1000$ times for arbitrary input states showed no failures, which provided strong motivation to use \Eq{30} and \Eq{32} as part of the ansatz for $\rhoX$ in \Eq{5} (which was then \textit{proven} to be a full solution, as shown in \Sec{III.A}).}
\label{fig:3}
\end{figure}

As \Fig{3} shows, $\alpha$ and $\beta$ do \textit{not} generally have the same roles (due to the ordered eigenvalues). Furthermore, all states tested with $C>0$ had the property that $C(\alpha ,\beta )$ always intersected the $C$ plane at $\beta=0\text{ or }\frac{\pi}{2}$. This suggested that, in our ansatz, we may be able to set 
\begin{Equation}                      {30}
\beta=0,
\end{Equation}
which, when put into \Eq{29} and setting $C(\alpha,0)=C$, gives
\begin{Equation}                      {31}
C = \max \{ 0,(\lambda _1  - \lambda _3 )s_{2\alpha }  - 2\sqrt {\lambda _2 \lambda _4 } \}.
\end{Equation}
Solving \Eq{31} for $\alpha$ when $(\lambda _1  \hsp{-2}-\hsp{-2} \lambda _3 )s_{2\alpha }  \hsp{-2}-\hsp{-2} 2\sqrt {\lambda _2 \lambda _4 }\geqslant 0$ yields
\begin{Equation}                      {32}
\alpha  = \left\{ {\begin{array}{*{20}l}
   {\frac{1}{2}\sin ^{ - 1} (\frac{{C + 2\sqrt {\lambda _2 \lambda _4 } }}{{\lambda _1  - \lambda _3 }});} & {\lambda _1  \ne \lambda _3 }  \\
   {\frac{\pi }{4};\rule{0pt}{9pt}} & {\lambda _1  = \lambda _3 ,}  \\
\end{array}} \right.
\end{Equation}
where the $\lambda _1 \hsp{-1} =\hsp{-1} \lambda _3$ case was found by observing that since $\lambda_1 \hsp{-1}\geqslant\hsp{-1}\cdots\hsp{-1}\geqslant\hsp{-1}\lambda_4 \hsp{-1}\geqslant\hsp{-1} 0$ and $\sum\nolimits_{k=1}^{4}{\lambda_k}\hsp{-1}=\hsp{-1}1$, then $\lambda _1  \hsp{-1}=\hsp{-1} \lambda _3$ and $(\lambda _1  \hsp{-1}-\hsp{-1} \lambda _3 )s_{2\alpha }  \hsp{-1}-\hsp{-1} 2\sqrt {\lambda _2 \lambda _4 }\hsp{-1}\geqslant \hsp{-1}0$ imply that $\lambda _1  \hsp{-1}=\hsp{-1} \lambda _2 \hsp{-1} =\hsp{-1} \lambda _3  \hsp{-1}=\hsp{-1} \frac{1}{3}$ and $\lambda _4 \hsp{-1}=\hsp{-1}0$, which causes $\lambda _1  \hsp{-1}-\hsp{-1} \lambda _3  \hsp{-1}-\hsp{-1} 2\sqrt {\lambda _2 \lambda _4 } \hsp{-1} =\hsp{-1} 0$, which by \Eq{14} and \Eq{16} implies $C\hsp{-1}=\hsp{-1}0$.  Since $\alpha$ is free when $\lambda _1 \hsp{-1} = \hsp{-1}\lambda _3$ and $(\lambda _1  \hsp{-1}-\hsp{-1} \lambda _3 )s_{2\alpha }  \hsp{-1}-\hsp{-1} 2\sqrt {\lambda _2 \lambda _4 }\hsp{-1}\geqslant \hsp{-1}0$, then for continuity, we can put $C\hsp{-1}=\hsp{-1}0$ and $2\sqrt {\lambda _2 \lambda _4 }=\lambda _1  \hsp{-1}-\hsp{-1} \lambda _3$ \textit{in the $\lambda _1  \hsp{-1}\ne\hsp{-1} \lambda _3$ case solution} to get \smash{$\alpha\hsp{-1}=\hsp{-1}\frac{1}{2}\sin ^{ - 1} (\frac{{\lambda _1  - \lambda _3 }}{{\lambda _1  - \lambda _3 }}) \hsp{-1}=\hsp{-1} \frac{\pi }{4}$} for the $\lambda _1 \hsp{-1} = \hsp{-1}\lambda _3$ case by using l'H{\^o}pital's rule on \smash{$\mathop {\lim }\limits_{x \to 0}(s_{2\alpha})$} with \smash{$x\hsp{-1}\equiv\hsp{-1}\lambda _1 \hsp{-1} -\hsp{-1} \lambda _3$}.
\subsubsection{\label{sec:III.B.3}Applying Parameters to Get $\rhoX$}
Now we simply need to plug our values of $\alpha$ and $\beta$ from \Eq{32} and \Eq{30} into \Eq{28} to get $\rhoX$ in terms of $C$ as well as spectrum. Solving \Eq{32} for $s_{2\alpha}$ and putting that and \Eq{30} into \Eq{28} with the help of the identity \smash{$s_\theta ^2 = $} \smash{$ \frac{1}{2}(1 - \sqrt {1 - s_{2\theta }^2 } )$}\rule{0pt}{9.1pt} for \smash{$\theta  \in [0,\frac{\pi }{2}]$}, we obtain
\begin{Equation}                      {33}
\rhoX\hsp{-1}=\hsp{-1}\frac{1}{2}\hsp{-3}\left(\hsp{-2} {\begin{array}{*{20}c}
   {\lambda _1  \hsp{-2}+\hsp{-2} \lambda _3  \hsp{-2}+\hsp{-2} \sqrt{Q\rule{0pt}{8pt}}} &\hsp{-2}  \cdot  &\hsp{-2}  \cdot  &\hsp{-2} {C \hsp{-2}+\hsp{-2} 2\sqrt {\lambda _2 \lambda _4 \rule{0pt}{8pt}} }  \\
    \cdot  &\hsp{-2} {2\lambda _2 } &\hsp{-2}  \cdot  &\hsp{-2}  \cdot   \\
    \cdot  &\hsp{-2}  \cdot  &\hsp{-2} {2\lambda _4 } &\hsp{-2}  \cdot   \\
   {C \hsp{-2}+\hsp{-2} 2\sqrt {\lambda _2 \lambda _4 \rule{0pt}{8pt}} } &\hsp{-2}  \cdot  &\hsp{-2}  \cdot  &\hsp{-2} {\lambda _1  \hsp{-2}+\hsp{-2} \lambda _3  \hsp{-2}-\hsp{-2} \sqrt{Q\rule{0pt}{8pt}}}  \\
\end{array}}\hsp{-1} \right)\hsp{-1},
\end{Equation}
where the cases of \Eq{32} united in this process, and
\begin{Equation}                      {34}
Q \equiv (\lambda _1  - \lambda _3 )^2  - (C + 2\sqrt {\lambda _2 \lambda _4 } )^2 .
\end{Equation}
However, the physicality of $\rhoX$ requires that the radicand $Q$ in \Eq{33} must be nonnegative, so we get the condition that \textit{\Eq{33} and thus \Eq{32} only apply when $Q\geqslant 0$}.

So what happens when $Q<0$? In fact, the answer lies in the proof in \hyperlink{LambdaCExhaustive:ii.b}{(ii.b)} of \Sec{III.A}; since $Q<0$ implies that $\lambda _1  - \lambda _3-2\sqrt {\lambda _2 \lambda _4 }<0$ which means that all MEMS with spectra that satisfy $\lambda _1  - \lambda _3-2\sqrt {\lambda _2 \lambda _4 }<0$ have $C_{\text{MEMS}_\Lambda}=0$, then \textit{all} states $\rho$ must have $C=0$ whenever $Q<0$, as concluded in \Eq{22}.

Thus, we are free to use \textit{any} separable X state of the same spectrum for the $Q<0$ case, so for the sake of continuity with the $Q\geqslant 0$ case, we chose the MEMS X state of \Eq{13}, since as explained above, its concurrence is zero whenever $\lambda _1  - \lambda _3-2\sqrt {\lambda _2 \lambda _4 }<0$, which is guaranteed when $Q<0$.  The continuity of the off-diagonal elements between $Q$ cases in \Eq{8} is preserved since in the $Q=0$ case, $C=\lambda _1  - \lambda _3-2\sqrt {\lambda _2 \lambda _4 }$ (put $Q=0$ into \Eq{34}), so $C+2\sqrt {\lambda _2 \lambda _4 }=\lambda _1  - \lambda _3$, showing that the off-diagonal elements have equivalent values at the edge case. 

Therefore, all of this motivates the form of $\rhoX$ in \Eq{8}.  Then, the $Q$ cases can be united by defining 
\begin{Equation}                      {35}
\Omega\equiv\max\{0,Q\},
\end{Equation}
which allows us to replace $\sqrt{Q}$ in \Eq{8} (or \Eq{33}) with $\sqrt{\Omega}$ since that causes no change when $Q\geqslant 0$, and allows us to \textit{add} $\pm\sqrt{\Omega}$ to ${\rhoX}_{1,1}$ and ${\rhoX}_{4,4}$ of the $Q<0$ case in \Eq{8} since that is merely adding zero.  The off-diagonal elements are unified by solving \Eq{34} for $C+2\sqrt {\lambda _2 \lambda _4 }$ as
\begin{Equation}                      {36}
C+2\sqrt {\lambda _2 \lambda _4 }=\sqrt{(\lambda _1  - \lambda _3 )^2 -Q},
\end{Equation}
and then replacing $Q$ by $\Omega$ as
\begin{Equation}                      {37}
C+2\sqrt {\lambda _2 \lambda _4 }=\sqrt{(\lambda _1  - \lambda _3 )^2 -\Omega},
\end{Equation}
since in the $Q\geqslant 0$ case, $\Omega=Q$ which makes \Eq{37} agree with \Eq{36} and can be directly substituted into \Eq{5} to get back the $Q\geqslant 0$ case in \Eq{8}, while in the $Q<0$ case, $\Omega=0$ which makes \Eq{37} simplify to $\lambda _1  - \lambda _3$ which causes \Eq{5} to simplify to the $Q<0$ case of \Eq{8}.

Therefore, we have now shown how the state family $\rhoX$ in \Eq{5} was derived, and more importantly, we showed in \Sec{III.A} an explicit \textit{proof} that $\rhoX$ in \Eq{5} does indeed have EPU equivalence with all general states $\rho$.  The explanations in the present section are merely the preliminary work for proposing a candidate family of X states as an ansatz, while the proofs in \Sec{III.A} conclusively verify that $\rhoX$ in \Eq{5} is necessary and sufficient to achieve universal EPU equivalence.
\section{\label{sec:IV}Conclusions}
The main result of this paper is the explicit family of X states $\rhoX$ given in \Eq{5} with the property of entanglement-preserving unitary (EPU) equivalence to the set of general two-qubit states $\rho$. This means that for every $\rho$, it is possible to find a \textit{unitary} operation that converts $\rho$ to an X state of the same entanglement as $\rho$.

The family $\rhoX$ in \Eq{5} gives us the explicit \textit{result} of such an EPU transformation, given only the spectrum of $\rho$ and its concurrence $C$.  The unitary transformation itself can then be found implicitly from \Eq{7}, though it is unnecessary since \Eq{5} gives the desired result. Explicit solution of the EPU is possible, but not nearly as practical as \Eq{5}.

The \textit{proof} that $\rhoX$ of \Eq{5} is EPU-equivalent to general states was given in \Sec{III.A}, while the \textit{motivation} for using \Eq{5} as an ansatz prior to proof was given in \Sec{III.B}. This method of finding a useful ansatz may be helpful for investigating EPU equivalence in larger systems.

It was already proven in \cite[]{MeMH}, in agreement with \cite[]{HedX}, that in qubit-qutrit ($2\times 3$) systems, literal X states cannot achieve EPU equivalence to general states. However, the more general special state family called true-generalized X (TGX) states proposed in \cite[]{HedX} may indeed have EPU equivalence with general states, though a proof for that is still unknown for $2\times 3$ or larger systems.

A particularly useful application of \Eq{5} is that it lets us \textit{parameterize mixed states of any spectrum and concurrence}; that is, given some desired spectrum and $C$, plug them into \Eq{5} to get a state with the desired properties.  However, for this purpose, the \textit{physical} values of $C$ are limited by the eigenvalues as shown in \Eq{14} and \Eq{16}, so for a free parameterization, we can rewrite \Eq{6} as 
\begin{Equation}                      {38}
Q \equiv (\lambda _1  - \lambda _3 )^2  - (C_{\eta} + 2\sqrt {\lambda _2 \lambda _4 } )^2 ,
\end{Equation}
where $C_{\eta}$ is the \textit{guaranteed-physical concurrence},
\begin{Equation}                      {39}
C_{\eta}\equiv \eta\max\{0,\lambda _1  - \lambda _3 -2\sqrt {\lambda _2 \lambda _4 }\},
\end{Equation}
where $\eta\in[0,1]$ is free, and \Eq{39} ensures that we always generate $C\in[0,\max\{0,\lambda _1  - \lambda _3 -2\sqrt {\lambda _2 \lambda _4 }\}]$. Then, by using \Eq{38} in \Eq{5}, we can choose any combination of spectrum (parameterizable as squared hyperspherical coordinates \cite[]{HedU} $\text{s.t.}\,\lambda_{1}\geqslant\cdots\geqslant\lambda_{4}$) and $\eta$ (which scales the physically achievable entanglement), and be guaranteed to generate a physical $\rhoX$. Alternatively, just choose $C\in[0,\max\{0,\lambda _1  - \lambda _3 -2\sqrt {\lambda _2 \lambda _4 }\}]$.  (Note that this is not a concern in the other direction, since given a general $\rho$, its physicality guarantees that its spectrum and $C$ are a physically valid combination to use in \Eq{5} and \Eq{6}.)

One area of future research is to find a \textit{general form} of the \smash{$U_{{\text{EPU}}_\text{X}}$} connecting $\rhoX$ of \Eq{5} to general $\rho$.  As shown in \cite[]{HedE}, for two qubits, \textit{pure} X states are EPU equivalent to all pure general states by an EPU of the form 
\begin{Equation}                      {40}
U_{{\text{EPU}}}=(U^{(1)}\otimes U^{(2)})D,
\end{Equation}
where $U^{(m)}$ for $m=1,2$ are unitary operators on the local subsystems, but $D$ is any diagonal unitary operator, and is not necessarily local, as proven in \cite[]{HedX}, and $D$ must be adjacent to $\rhoX$, such as \smash{$\rho=U_{{\text{EPU}}}\rhoX U_{{\text{EPU}}}^{\dag}$}. However, for \textit{mixed} two-qubit states, there may be a more general form for \smash{$U_{{\text{EPU}}_\text{X}}^{\dag}$} (we speak of the adjoint of \Eq{7} because here we are applying it in the opposite direction as a parameterization of general states rather than a simplifying transformation away from them as it was defined, while \Eq{40} is defined in the parameterization context).  Thus, while this paper proves that it is possible to parameterize \textit{all} two-qubit states $\rho$ in terms of spectrum and concurrence as, inversely to \Fig{1},
\vspace{-4pt}
\begin{Equation}                      {41}
\rho=U_{{\text{EPU}}_\text{X}}^{\dag}\rhoX U_{{\text{EPU}}_\text{X}},
\vspace{-4pt}
\end{Equation}
for some general \smash{$U_{{\text{EPU}}_\text{X}}$}, we can only \textit{hypothesize} that \smash{$U_{{\text{EPU}}}^{\dag}$} of \Eq{40} is the form of that general \smash{$U_{{\text{EPU}}_\text{X}}$} at this time.  One reason that \Eq{40} may \textit{not} be the general EPU we seek is that technically, the form of $U_{{\text{EPU}}_\text{X}}$ is already specified in \Eq{7}, but \Eq{7}  generally depends on the eigenvalues and concurrence when written fully explicitly due to its construction from eigenvectors, while \Eq{40} can be chosen completely independent of spectrum or concurrence.  It may be that such dependencies \textit{cancel} in \Eq{7}, resulting in the adjoint of \Eq{40}, but at the present time, the complexity of the problem prevents a clear answer.

Interestingly, \smash{$\rhoX$} of \Eq{5} is only a \textit{subset} of X states. Taking advantage of this fact, if we apply the local EPU \smash{$U_{\text{L}}\equiv\binom{0\hsp{3}1}{1\hsp{3}0}\otimes\binom{1\hsp{3}0}{0\hsp{3}1}$}\rule{0pt}{9.6pt} to $\rhoX$ to get \smash{$\rhoX'\equiv U_{\text{L}}\rhoX U_{\text{L}}^{\dag}$} as
\vspace{-2pt}
\begin{Equation}                      {42}
\rhoX'\equiv\frac{1}{2}\hsp{-2}\left(\hsp{-0.5} {\begin{array}{*{20}c}
   {2\lambda _4} &\hsp{-4}  \cdot  &\hsp{-2}  \cdot  &\hsp{-4} \cdot  \\
    \cdot  &\hsp{-4} {\lambda _1  \hsp{-1}+\hsp{-1} \lambda _3  \hsp{-1}-\hsp{-1} \sqrt {\Omega\rule{0pt}{8.2pt}} } &\hsp{-2} {\sqrt {(\lambda _1  \hsp{-1}-\hsp{-1} \lambda _3 )^2  \hsp{-1}-\hsp{-1} \Omega\rule{0pt}{8.2pt}} }  &\hsp{-4}  \cdot   \\
    \cdot  &\hsp{-4} {\sqrt {(\lambda _1  \hsp{-1}-\hsp{-1} \lambda _3 )^2  \hsp{-1}-\hsp{-1} \Omega\rule{0pt}{8.2pt}} } &\hsp{-2} {\lambda _1  \hsp{-1}+\hsp{-1} \lambda _3  \hsp{-1}+\hsp{-1} \sqrt {\Omega\rule{0pt}{8.2pt}}} &\hsp{-4}  \cdot   \\
    \cdot  &\hsp{-4}  \cdot  &\hsp{-2}  \cdot  &\hsp{-4} {2\lambda _2}  \\
\end{array}}\hsp{-0.5} \right)\hsp{-1},
\end{Equation}
where $\Omega$ is defined as in \Eq{5}, it may then be \textit{more robust} against certain types of noise \cite[]{YuEb}, and its general EPU equivalence means that \textit{any} intended $\rho$ can be unitarily converted to such an initial form, if $\rho$ is known.

In closing, the state family $\rhoX$ of \Eq{5} (and $\rhoX'$ of \Eq{42}) provides an explicit result that proves the conjecture of \cite[]{HedX}, and also validates the proof of the existence of EPU transformations from \cite[]{MeMG}. These states are likely to have a wide range of useful applications in both technology and theoretical work, and it is especially hoped that they will give us insight to entanglement in larger systems, where similar simple forms may help us find computable entanglement measures for mixed states.
\vspace{-4pt}
\begin{acknowledgments}
\vspace{-4pt}
Many thanks to Ting Yu for helpful discussions.
\end{acknowledgments}
%
\end{document}